\documentclass[11pt]{article}
\usepackage{amsmath}
\usepackage{amsfonts}
\usepackage{amssymb}
\usepackage{graphicx}
\usepackage{bm}
\usepackage{color}
\usepackage{amscd}

\def\bea{\begin{eqnarray}}
\def\eea{\end{eqnarray}}

\begin{document}
\begin{center}
\LARGE {\bf Near Horizon Symmetries of the Non-Extremal Black Hole Solutions of Generalized Minimal Massive Gravity }
\end{center}

\begin{center}
{M. R. Setare \footnote{E-mail: rezakord@ipm.ir}\hspace{1mm} ,
H. Adami \footnote{E-mail: hamed.adami@yahoo.com}\hspace{1.5mm} \\
{\small {\em  Department of Science, University of Kurdistan, Sanandaj, Iran.}}}\\

\end{center}

\begin{center}
{\bf{Abstract}}\\
We consider the Generalized Minimal Massive Gravity (GMMG) model in the first order formalism. We show that all the solutions of the Einstein gravity with negative cosmological constants solve the equations of motion of considered model. Then we find an expression for the off-shell conserved charges of this model. By considering the near horizon geometry of a three dimensional black hole in the Gaussian null coordinates, we find near horizon conserved charges and their algebra. The obtained algebra is centrally extended. By writing the algebra of conserved charges in terms of Fourier modes and considering the BTZ black hole solution as an example, one can see that the charge
associated with rotations along $\mathcal{Y}_{0}$ coincides exactly
with the angular momentum, and he charge associated with time translations $\mathcal{T}_{0}$ is the product of the black hole entropy and its temperature. As we expect, in the limit when the GMMG tends to the Einstein gravity, all the result we obtain in this paper reduce to the result of the paper \cite{5}.  
\end{center}

\section{Introduction}
It is well known that Topologically Massive Gravity (TMG)\cite{4'} and New theory of Massive Gravity (NMG) \cite{2'} in three dimensions have a bulk-boundary
unitarity conflict. Either the bulk or the boundary theory is non-unitary, so there is a clash between the positivity of the two Brown-Henneaux boundary $c$ charges and the bulk energies. Recently the authors of \cite{3'} proposed a new model, named Minimal Massive Gravity (MMG), which has the same minimal local structure as TMG. The MMG model has the same gravitational
degree of freedom as the TMG has and the linearization of the metric field equations for MMG yield a single propagating
massive spin-2 field. It seems that the single massive degree of freedom of MMG is
unitary in the bulk and gives rise to a unitary CFT on the boundary. Following this paper some interesting works have been done on MMG model \cite{18} (see also \cite{4,10}). More recently the first author has introduced  Generalized Minimal Massive Gravity (GMMG) \cite{2}. GMMG is realized by adding higher-derivative deformation term to the Lagrangian of MMG. As has been shown in \cite{2}, GMMG also avoids the aforementioned ``bulk-boundary unitarity clash''. Hamiltonian analysis show that the GMMG model has no Boulware-Deser ghosts and this model propagate only two physical modes.
 So this model is viable candidate for semi-classical limit of a unitary quantum $3D$ massive gravity.\\
 Recently Donnay et al \cite{5}, have shown that the asymptotic symmetries close to the horizon of the nonextremal black hole solution of the three-dimensional Einstein gravity in the presence of
a negative cosmological term, are generated
by an extension of supertranslations. The near
horizon symmetries in three dimensions are related with the Bondi–-van der Burg–-Metzner–-Sachs (BMS) algebra \cite{5'}. The authors of \cite{5} have shown that for a special choice of boundary conditions, the
near region to the horizon of a stationary black hole
present a generalization of supertranslation, including a
semidirect sum with superrotations, represented by Virasoro
algebra. To see about the importance of near-horizon BMS symmetries as a means
to understand black holes refer to recent paper by Hawking et al \cite{6'}. For another related works see \cite{7'}.\\
In this paper we are going to generalize the work of \cite{5} from Einstein gravity to the GMMG model. Since it seems that this model is viable candidate for semi-classical limit of a unitary quantum gravity in three-dimension, the requested extension is an interesting task. At first we show that  the solutions of the Einstein gravity with negative cosmological constant are solutions of GMMG also. Then we find an expression for off-shell conserved charge of GMMG associated with an arbitrary vector field $\xi$.
 We will show that in the context of GMMG model the algebra which have obtained in \cite{5} for the near horizon geometry of black hole solutions in the cosmological Einstein gravity (the Einstein gravity in the presence of negative cosmological constant) will be centrally extended. Then we will see that the zero mode charge $\mathcal{T}_{0}$ (the charge associated with time translations) is proportional to the entropy of the BTZ black hole solution of GMMG,  and $\mathcal{Y}_{0}$ (the charge
associated with rotations) gives us the angular momentum of the BTZ black hole.
\section{The Generalized Minimal Massive Gravity}
In this paper, we work in the first order formalism. In the first order formalism dreibein $e^{a}=e^{a}_{\hspace{1.5 mm} \mu} dx^{\mu}$ and dualized spin-connection $\omega ^{a}= \omega ^{a}_{\hspace{1.5 mm} \mu} dx^{\mu}$ both are treated as independent Lorentz vector valued 1-form fields. We use Latin and Greek Letters to characterize Lorentz and coordinate indices, respectively. Also, we use a 3D-vector algebra notation for Lorentz vectors (see
for instance \cite{1}). The Lagrangian of Generalized Minimal Massive Gravity (GMMG) is given by \cite{2}
\begin{equation}\label{1}
\begin{split}
   L= & - \sigma e \cdot R(\omega) + \frac{\Lambda _{0}}{6} e \cdot e \times e + \frac{1}{2 \mu} \left( \omega \cdot d \omega + \frac{1}{3} \omega \cdot \omega \times \omega \right) \\
     & - \frac{1}{m^{2}} \left( f \cdot R + \frac{1}{2} e \cdot f \times f \right) + h \cdot T(\omega) + \frac{\alpha}{2} e \cdot h \times h,
\end{split}
\end{equation}
where $R(\omega)$ and $T(\omega)$ are dualized curvature and torsion 2-forms
\begin{equation}\label{2}
  R=d \omega + \frac{1}{2} \omega \times \omega , \hspace{1.5 cm} T = D(\omega) e = de+ \omega \times e,
\end{equation}
respectively. In the above Lagrangian, $\sigma$, $\Lambda _{0}$, $\mu$, $m$ and $\alpha$ are a sign, cosmological parameter with dimension of mass squared, mass parameter of Lorentz Chern-Simons term, mass parameter of New Massive Gravity term and a dimensionless parameter, respectively. Also, $f^{a}$ and $h^{a}$ are Lorentz vector valued 1-form auxiliary fields.\\
An arbitrary variation of the Lagrangian \eqref{1} is given by
\begin{equation}\label{3}
     \delta L = \delta e \cdot E _{e} + \delta \omega \cdot E _{\omega} + \delta f \cdot E _{f} + \delta h \cdot E _{h} + d \Theta (\Phi ,  \delta \Phi),
\end{equation}
where $\Phi$ is a collection of all the fields, i.e. $\Phi = \{ e, \omega , f , h \}$. In the above equation, we have the following definitions
\begin{equation}\label{4}
  E_{e} = - \sigma R (\omega) + \frac{\Lambda _{0}}{2} e \times e + D(\omega) h - \frac{1}{2 m^{2}} f \times f  + \frac{\alpha}{2} h \times h  ,
\end{equation}
\begin{equation}\label{5}
  E_{\omega} =- \sigma T(\omega) + \frac{1}{\mu} R(\omega) - \frac{1}{m^{2}} D (\omega) f + e \times h ,
\end{equation}
\begin{equation}\label{6}
  E_{f} = - \frac{1}{m^{2}} \left( R(\omega) + e \times f \right) ,
\end{equation}
\begin{equation}\label{7}
  E_{h} = T(\omega) + \alpha e \times h ,
\end{equation}
\begin{equation}\label{8}
  \Theta (\Phi , \delta \Phi) = - \sigma \delta \omega \cdot e + \frac{1}{2 \mu} \delta \omega \cdot \omega - \frac{1}{m^{2}} \delta \omega \cdot f + \delta e \cdot h.
\end{equation}
The equations of motion of  GMMG are
\begin{equation}\label{9}
 E _{e} = E _{\omega} = E _{f} = E _{h} =0,
\end{equation}
and $\Theta (\Phi , \delta \Phi)$ is just surface term. The equations of motion of the considered theory can be rewritten as
\begin{equation}\label{10}
   - \sigma R (\Omega) + (1 + \sigma \alpha ) D(\Omega) h - \frac{1}{2} \alpha (1 + \sigma \alpha ) h \times h + \frac{\Lambda _{0}}{2} e \times e - \frac{1}{2 m^{2}} f \times f  =0,
\end{equation}
\begin{equation}\label{11}
  - e \times f + \mu (1 + \sigma \alpha ) e \times h - \frac{\mu}{m^{2}} D(\Omega) f + \frac{\mu \alpha}{m^{2}} h \times f=0 ,
\end{equation}
\begin{equation}\label{12}
  R(\Omega) - \alpha D(\Omega) h + \frac{1}{2} \alpha ^{2} h \times h + e \times f =0,
\end{equation}
\begin{equation}\label{13}
  T(\Omega) = 0 .
\end{equation}
where $\Omega = \omega - \alpha h$ is ordinary torsion-free dualized spin-connection.\\
For the solutions of the Einstein-Hilbert gravity with negative cosmological constant we have
\begin{equation}\label{14}
  R(\Omega)+\frac{1}{2 l^{2}} e \times e =0,
\end{equation}
where $l$ denotes AdS radius. In the context of the GMMG we consider following ansatz
\begin{equation}\label{15}
  f^{a}=Fe^{a}, \hspace{1 cm} h^{a}=He^{a},
\end{equation}
where $F$ and $H$ are constant parameters. By substituting Eq.\eqref{14} and Eq.\eqref{15} into the equations of motion \eqref{10}-\eqref{13} we obtain
\begin{equation}\label{16}
  \frac{\sigma}{ l^{2}} - \alpha (1 + \sigma \alpha ) H ^{2} + \Lambda _{0} - \frac{F^{2}}{ m^{2}}=0,
\end{equation}
\begin{equation}\label{17}
  - \frac{1}{\mu l^{2}} + 2 (1 + \sigma \alpha ) H + \frac{2 \alpha}{m^{2}} F H + \frac{\alpha ^{2}}{\mu} H^{2}=0,
\end{equation}
\begin{equation}\label{18}
  - F + \mu (1 + \sigma \alpha ) H + \frac{\mu \alpha}{m^{2}} FH=0.
\end{equation}
Thus, solutions of the Einstein-Hilbert gravity with negative cosmological constant are solutions of GMMG if $\Lambda _{0}$, $F$ and $H$ satisfy equations \eqref{16}-\eqref{18}.
\section{Off-shell Conserved Charges}
In this section, we find an expression for off-shell conserved charge of GMMG associated with an arbitrary vector field $\xi$.\\
Under Lorentz gauge transformation $\Lambda \in SO(2,1) $ dreibein transforms as $e ^{a}_{\hspace{1.5 mm} \mu} \rightarrow \Lambda ^{a}_{\hspace{1.5 mm} b} e^{b}_{\hspace{1.5 mm} \mu}$ so that the spacetime metric $g_{\mu \nu}=\eta _{ab} e^{a}_{\hspace{1.5 mm} \mu} e^{b}_{\hspace{1.5 mm} \nu}$ under this transformation remains unchanged. Also, under Lorentz gauge transformation the spin-connection transforms as $\omega \rightarrow \Lambda \omega \Lambda ^{-1}+ \Lambda d \Lambda ^{-1}$ so this is not an invariant quantity under considered transformation. One can define Lorentz-Lie (L-L) derivative of the dreibein 1-form as \cite{3}
\begin{equation}\label{19}
  \mathfrak{L}_{\xi} e^{a} = \pounds_{\xi} e^{a} +\lambda ^{a}_{\hspace{1.5 mm} b} e^{b}.
\end{equation}
where $\pounds_{\xi}$ denotes ordinary Lie derivative along $\xi$ and $\lambda ^{a}_{\hspace{1.5 mm} b}$ generates the Lorentz gauge transformations $SO(2,1)$. In general, $\lambda ^{a}_{\hspace{1.5 mm} b}$ is independent of the dynamical fields of considered model and it is a function of spacetime coordinates and of the diffeomorphism generator $\xi$. The total variation of the dreibein and the spin-connection are defined as \cite{4}
\begin{equation}\label{20}
  \delta _{\xi} e^{a} = \mathfrak{L}_{\xi} e^{a},
\end{equation}
 \begin{equation}\label{21}
   \delta _{\xi} \omega = \mathfrak{L}_{\xi} \omega -d \chi _{\xi},
 \end{equation}
respectively, where $\chi^{a} _{\xi} = \frac{1}{2} \varepsilon ^{a}_{\hspace{1.5 mm} bc} \lambda ^{bc}$. The extra term in \eqref{21}, $-d \chi _{\xi}$, can makes a theory non-covariant, in the meaning of Lorentz covariance (Because $e$ and $\omega$ both are invariant under general coordinate transformations).\\
Now, we suppose that the variation of Lagrangian \eqref{3} is due to a diffeomorphism which is generated by the vector field $\xi$, then the total variation of Lagrangian \eqref{3} with respect to the diffeomorphism $\xi$ is
\begin{equation}\label{22}
    \delta _{\xi} L = \delta _{\xi} e \cdot E _{e} + \delta _{\xi} \omega \cdot E _{\omega} + \delta _{\xi} f \cdot E _{f} + \delta _{\xi} h \cdot E _{h} + d \Theta (\Phi ,  \delta _{\xi} \Phi).
\end{equation}
The presence of Lorentz Chern-Simons term in the Lagrangian \eqref{1} makes GMMG to be Lorentz non-covariant, by virtue of Eq.\eqref{21}. So, the total variation of the Lagrangian \eqref{1} due to diffeomorphism generator $\xi$ can be written as
\begin{equation}\label{23}
  \delta _{\xi} L = \mathfrak{L}_{\xi} L +  d \psi _{\xi}.
\end{equation}
From definition of total variation due to $\xi$, equations \eqref{20} and \eqref{21}, we can write
\begin{equation}\label{24}
  \delta _{\xi} e = D(\omega) i_{\xi} e + i_{\xi} T(\omega) + (\chi _{\xi} -  i_{\xi} \omega) \times e ,
\end{equation}
\begin{equation}\label{25}
  \delta _{\xi} \omega =  i_{\xi} R(\omega) + D(\omega) ( i_{\xi} \omega - \chi _{\xi}) ,
\end{equation}
\begin{equation}\label{26}
  \delta _{\xi} f = D(\omega) i_{\xi} f + i_{\xi} D(\omega) f + (\chi _{\xi} -  i_{\xi} \omega) \times f ,
\end{equation}
\begin{equation}\label{27}
  \delta _{\xi} h = D(\omega) i_{\xi} h + i_{\xi} D(\omega) h + (\chi _{\xi} -  i_{\xi} \omega) \times h ,
\end{equation}
where $i_{\xi}$ denotes the interior product in $\xi$. By substituting equations \eqref{23}-\eqref{27} into Eq.\eqref{22} we have
\begin{equation}\label{28}
\begin{split}
     d J (\xi) = & (i_{\xi} \omega - \chi _{\xi} ) \cdot \left[ D(\omega) E_{\omega} + e \times E_{e} + f \times E_{f} + h \times E_{h} \right]\\
     & + i_{\xi} e \cdot D(\omega) E_{e} + i_{\xi} f \cdot D(\omega) E_{f} + i_{\xi} h \cdot D(\omega) E_{h} \\
     & - i_{\xi} T(\omega) \cdot E_{e} - i_{\xi} R(\omega) \cdot E_{\omega} - i_{\xi} D(\omega) f \cdot E_{f} - i_{\xi} D(\omega) h \cdot E_{h} ,
\end{split}
\end{equation}
where
\begin{equation}\label{29}
  J(\xi) = \Theta (\Phi ,  \delta _{\xi} \Phi) - i_{\xi} L - \psi _{\xi} + i_{\xi} e \cdot E _{e} + (i_{\xi} \omega - \chi _{\xi} ) \cdot E _{\omega} + i_{\xi} f \cdot E _{f}+ i_{\xi} h \cdot E _{h}.
\end{equation}
The right hand side of the equation \eqref{28} vanishes by virtue of the Bianchi identities
\begin{equation}\label{30}
  D(\omega) R(\omega) = 0 , \hspace{1.5 cm} D(\omega) T(\omega) = R (\omega) \times e.
\end{equation}
Therefore, we find that
\begin{equation}\label{31}
  d J (\xi) =0.
\end{equation}
So $J (\xi)$ is an off-shell conserved current associated with an arbitrary vector field $\xi$. It is straightforward to calculate $\psi _{\xi}$ in Eq.\eqref{23}
\begin{equation}\label{32}
  \psi _{\xi} = \frac{1}{2 \mu} d \chi _{\xi} \cdot \omega,
\end{equation}
to obtain above equation we used the fact that exterior derivative and L-L derivative do not commute
\begin{equation}\label{33}
  \left[ d , \mathfrak{L}_{\xi} \right] e = d \chi _{\xi} \times e.
\end{equation}
Since $J(\xi)$ is closed then it is exact by virtue of the Poincare lemma. Thus, we find that
\begin{equation}\label{34}
  J(\xi) =  d K (\xi),
\end{equation}
where
\begin{equation}\label{35}
     K(\xi) = (i_{\xi} \omega - \chi _{\xi}) \cdot \left( - \sigma e + \frac{1}{2 \mu} \omega - \frac{1}{m^{2}} f\right) + i_{\xi} e \cdot h .
\end{equation}
Now, we can define the off-shell conserved charge associated with a vector field $\xi$ as
\begin{equation}\label{36}
  Q(\xi) = \frac{1}{8 \pi G} \int_{\Sigma} K(\xi),
\end{equation}
where $G$ denotes Newtonian gravitational constant and $\Sigma $ is a space-like codimension two surface. One can simplify $K(\xi)$ for a class of solutions which were presented in the previous section (see equations \eqref{14} and \eqref{15})
\begin{equation}\label{37}
\begin{split}
   K(\xi)= & - \left( \sigma + \frac{\alpha H}{2 \mu} + \frac{F}{m^{2}}  \right) (i_{\xi} \Omega - \chi _{\xi}) \cdot e + \frac{1}{2 \mu} (i_{\xi} \Omega - \chi _{\xi}) \cdot \Omega \\
     & - \frac{\alpha H}{2 \mu} i_{\xi} e \cdot \Omega + \frac{1}{2 \mu l^{2}} i_{\xi} e \cdot e ,
\end{split}
\end{equation}
which is expressed in terms of ordinary torsion-free spin-connection.
\section{Near Horizon Symmetries of Non-Extremal Black Holes}
In this section we summarize some results of the paper \cite{5}. The near horizon geometry of a 3D black hole in the Gaussian null coordinates is given by the following dreibein
\begin{equation}\label{38}
  \begin{split}
       e^{0} = & \sqrt{-A+\frac{C^{2}}{R^{2}}} dv - \frac{B}{\sqrt{-A+\frac{C^{2}}{R^{2}}}} d \rho \\
       e^{1} = & \frac{B}{\sqrt{-A+\frac{C^{2}}{R^{2}}}} d \rho \\
       e^{2} = & \frac{C}{R} dv + R d \phi
  \end{split}
\end{equation}
where $v$, $\rho$ and $\phi$ are the retarded time, the radial distance to the horizon and the angular coordinate, respectively. In this coordinates the horizon of black hole is located at $\rho=0$ . The metric corresponds to dreibein \eqref{38} is
\begin{equation}\label{39}
  ds^{2}= A dv^{2} + 2 B dv d \rho + 2 C dv d \phi + R ^{2} d \phi ^{2}.
\end{equation}
We consider the case in which functions $A$, $B$, $C$ and $R$ obey the following fall-off conditions close to the horizon \cite{5}
\begin{equation}\label{40}
  \begin{split}
       & A = - 2 \kappa \rho + \mathcal{O} (\rho ^{2}), \hspace{0.7 cm} B= 1 + \mathcal{O} (\rho ^{2}) \\
       & C= \theta (\phi) \rho + \mathcal{O} (\rho ^{2}), \hspace{0.7 cm} R^{2}=  \gamma (\phi) ^{2} + \beta (\phi) \rho + \mathcal{O} (\rho ^{2}),
  \end{split}
\end{equation}
where $\kappa$ is the surface gravity of considered black hole. Also, we demand that $g_{\rho \rho}= \mathcal{O}(\rho ^{2})$ and $g_{\rho \phi}= \mathcal{O}(\rho ^{2})$. We should mention that the boundary conditions \eqref{40} break the Poincare symmetry. The near horizon Killing vectors are as \cite{5}
\begin{equation}\label{41}
  \begin{split}
     \xi ^{v} = & T(\phi) + \mathcal{O} (\rho ^{3}) \\
     \xi ^{\rho} = & \frac{\theta (\phi) T^{\prime} (\phi)}{2 \gamma (\phi)^{2}} \rho ^{2} + \mathcal{O} (\rho ^{3}) \\
     \xi ^{\phi} = & Y(\phi) - \frac{ T^{\prime} (\phi)}{ \gamma (\phi)^{2}} \rho + \frac{\beta (\phi) T^{\prime} (\phi)}{2 \gamma (\phi)^{4}} \rho ^{2}+ \mathcal{O} (\rho ^{3})
  \end{split}
\end{equation}
preserve the fall-off conditions \eqref{40}. $T(\phi)$ and $Y(\phi)$ are arbitrary functions, and the prime denotes differentiation with respect to $\phi$. Under transformation generated by the Killing vector fields \eqref{41} the arbitrary functions $\theta(\phi)$, $\gamma(\phi)$ and $\beta(\phi)$, which have appeared in metric, transform as
\begin{equation}\label{42}
  \begin{split}
       & \delta _{\xi} \theta = \left( \theta Y \right)^{\prime} - 2 \kappa T^{\prime}, \hspace{0.7 cm} \delta _{\xi} \gamma = \left( \gamma Y \right)^{\prime}, \\
       & \delta _{\xi} \beta = 2 Y ^{\prime} \beta + 2 T^{\prime} \theta + Y \beta ^{\prime} - 2 T ^{\prime \prime} + \frac{2 \gamma ^{\prime} T^{\prime} }{\gamma}.
  \end{split}
\end{equation}
We introduce a modified version of the Lie brackets \cite{6}
\begin{equation}\label{43}
  \left[ \xi _{1} , \xi _{2} \right] = \pounds _{\xi _{1}} \xi _{2} - \delta ^{(g)} _{\xi _{1}} \xi _{2} + \delta ^{(g)} _{\xi _{2}} \xi _{1}
\end{equation}
so that the algebra of the near horizon Killing vector fields is close. In the equation \eqref{43}, $\delta ^{(g)} _{\xi _{1}} \xi _{2}$ denotes the change induced in $\xi _{2}$ due to the variation of metric $\delta  _{\xi _{1}} g_{\mu \nu} = \pounds _{\xi _{1}} g_{\mu \nu}$ \cite{6}. Thus, we have
\begin{equation}\label{44}
   \left[ \xi (T_{1},Y_{1}) , \xi (T_{2},Y_{2}) \right] = \xi (T_{12},Y_{12}),
\end{equation}
where
\begin{equation}\label{45}
  T_{12} = Y_{1} T^{\prime}_{2}-Y_{2} T^{\prime}_{1}, \hspace{0.7 cm} Y_{12} = Y_{1} Y^{\prime}_{2}-Y_{2} Y^{\prime}_{1}.
\end{equation}
By introducing Fourier modes $T_{n} = \xi \left( e^ {in\phi} , 0 \right)$ and $Y_{n} = \xi \left( 0 , e^ {in\phi} \right)$, one can find that $T_{n}$ and $Y_{n}$ satisfy the following algebra
\begin{equation}\label{46}
  \begin{split}
        i  \left[ T_{m} , T_{n} \right] = & 0,\\
        i  \left[ Y_{m} , Y_{n} \right] = & (m-n) Y_{m+n},\\
        i  \left[ Y_{m} , T_{n} \right] = & - n T_{m+n}.
  \end{split}
\end{equation}
 $T_{n}$ and $Y_{n}$ are generators of supertranslation and superrotation respectively.
\section{Near Horizon Conserved Charges and their Algebra in the Context of GMMG}
In this section we consider a class of solutions which were presented in the section 3 (see equations \eqref{14} and \eqref{15}). The conserved charges of such solutions can be obtained by Eq.\eqref{36} and Eq.\eqref{37}.\\
By demanding that the Lie-Lorentz derivative of $e^{a}$ becomes zero explicitly when $\xi$ is a Killing vector field, we find the following expression for $\chi_{\xi}$ \cite{4,7}
\begin{equation}\label{47}
  \chi _{\xi} ^{a} = i_{\xi} \omega ^{a} + \frac{1}{2} \varepsilon ^{a}_{\hspace{1.5 mm} bc} e^{\nu b} (i_{\xi} T^{c})_{\nu} + \frac{1}{2} \varepsilon ^{a}_{\hspace{1.5 mm} bc} e^{b}_{\hspace{1.5 mm} \mu} e^{c}_{\hspace{1.5 mm} \nu} \nabla ^{\mu} \xi ^{\nu} .
\end{equation}
It has been shown that this expression can be rewritten as \cite{8}
\begin{equation}\label{48}
  i_{\xi} \Omega - \chi _{\xi} = - \frac{1}{2} e _{\hspace{1.5 mm} \mu} \times e_{\hspace{1.5 mm} \nu} \tilde{\nabla} ^{\mu} \xi ^{\nu} ,
\end{equation}
where $\tilde{\nabla}$ denotes covariant derivative with respect to Levi-Civita connection.\\
Now, we take the space-like codimension two surface $\Sigma$ to be a circle with radius of $\rho \rightarrow 0$. Thus, the equation \eqref{36} can be rewritten as
\begin{equation}\label{49}
  Q(\xi) = \frac{1}{8 \pi G} \lim _{\rho \rightarrow 0} \int_{0} ^{2 \pi} K _{\phi} d \phi,
\end{equation}
where $K(\xi)$ is given by Eq.\eqref{37}. By substituting Eq.\eqref{38}, Eq.\eqref{48} and Eq.\eqref{41} into Eq.\eqref{49} we find the conserved charge corresponds to the Killing vector \eqref{41} as follows:
\begin{equation}\label{50}
\begin{split}
Q(\xi) = - \frac{1}{16 \pi G} \int_{0} ^{2 \pi} d \phi \{ & \left( \sigma + \frac{\alpha H}{2 \mu} + \frac{F}{m^{2}}  \right) \gamma (\phi) \left[ 2 \kappa T(\phi) - \theta (\phi) Y(\phi) \right] \\
      & + \frac{1}{4 \mu} \theta (\phi) \left[ 2 \kappa T(\phi) - \theta (\phi) Y(\phi) \right] \\
      & -  \frac{\alpha H}{2 \mu} \gamma (\phi) \theta (\phi) Y(\phi) - \frac{1}{\mu l^{2}} \gamma (\phi) ^{2} Y(\phi) \}
\end{split}
\end{equation}
In the limit of $\mu \rightarrow \infty$ and $m \rightarrow \infty$ and by setting $\sigma = -1$, where the GMMG model reduce to the Einstein gravity in the presence of negative cosmological constant, the above conserved charge reduced to the result of \cite{5} exactly.\\
We know that the algebra of conserved charges can be written as \cite{9}
\begin{equation}\label{51}
  \left\{ Q(\xi _{1}) , Q(\xi _{2}) \right\} = Q \left(  \left[ \xi _{1} , \xi _{2} \right] \right) + \mathcal{C} \left( \xi _{1} , \xi _{2} \right)
\end{equation}
where $\mathcal{C} \left( \xi _{1} , \xi _{2} \right)$ is central extension term. Also, the left hand side of the equation \eqref{51} can be defined by
\begin{equation}\label{52}
  \left\{ Q(\xi _{1}) , Q(\xi _{2}) \right\}= \frac{1}{2} \left( \delta ^{(g)} _{\xi _{2}} Q(\xi _{1}) - \delta ^{(g)} _{\xi _{1}} Q(\xi _{2}) \right).
\end{equation}
Thus, for the obtained conserved charge \eqref{50}, we find that
\begin{multline}\label{53}
     \left\{ Q(\xi _{1}) , Q(\xi _{2}) \right\} = Q \left(  \left[ \xi _{1} , \xi _{2} \right] \right) \\
        + \frac{\kappa}{64 \pi \mu G} \int_{0}^{2 \pi} d \phi \left\{ 2 \kappa \left(T_{1} T^{\prime}_{2} - T_{2} T^{\prime}_{1} \right) - \left( \theta (\phi) + 2 \alpha H \gamma (\phi) \right) T_{12}\right\}.
\end{multline}
For Fourier modes ,$\mathcal{T}_{n} = Q(e^{in\phi},0) $ and $\mathcal{Y}_{n} = Q(0,e^{in\phi}) $, we have
\begin{equation}\label{54}
  \begin{split}
        i  \left[ \mathcal{T} _{m} , \mathcal{T}_{n} \right] = & - \frac{\kappa ^{2} n}{8 \mu G} \delta _{m+n,0},\\
        i  \left[ \mathcal{Y} _{m} , \mathcal{Y}_{n} \right] = & (m-n) \mathcal{Y}_{m+n},\\
        i  \left[ \mathcal{Y}_{m} , \mathcal{T}_{n} \right] = & - n \mathcal{T}_{m+n} + \frac{\kappa n}{64 \pi \mu G} \int_{0}^{2 \pi} e^{i(m+n)\phi} \left\{ \theta (\phi) + 2 \alpha H \gamma (\phi) \right\} d \phi.
  \end{split}
\end{equation}
In the limit of $\mu \rightarrow \infty$, this algebra reduced to the result that appeared in \cite{5}. In other words, the central extension term comes from just Lorentz Chern-Simons term and in the absence of this term we do not have central extension term. In the framework of the cosmological Einstein gravity, the algebra spanned by $\mathcal{T} _{m}$ and $\mathcal{Y}_{n}$ is isomorphic to Eq.\eqref{46}, with no central extensions \cite{5}. By looking at the experssion of conserved charge we obtained in Eq.\eqref{50}, and from the above algebra Eq.\eqref{54} it is clear that, to obtain the algebra of \cite{5} it is necessary only we turn-off the topological term in the Lagrangian of the GMMG, i.e $\mu \rightarrow \infty$. The presence of term proportional to the $\frac{1}{m^2}$ (the term of the NMG), does not lead to a centrally extended algebra. If we introduce following generator
\begin{equation}\label{550}
  \mathcal{P}_{n} =\sum_{k\in Z}\mathcal{T}_{k}\mathcal{T}_{n-k},
\end{equation}
simply one can show that the algebra spanned by $\mathcal{P}_{n}$ and $\mathcal{Y}_{n}$ is $BMS_3$ \cite{5,19}. So according to the above discussion we can claim that the near horizon
geometry of a non-extremal black hole solutions of NMG, similar to the same solutions of the Einstein gravity, have $BMS_3$ symmetry which can be recovered by means of the mentioned Sugawara construction. \\
 One can easily read off the eigenvalues of $\mathcal{T}_{n} $ and $\mathcal{Y}_{n} $ from \eqref{50}
\begin{equation}\label{55}
  \mathcal{T}_{n} = - \frac{\kappa}{8 \pi G} \int_{0}^{2 \pi} e^{in\phi} \left\{ \left( \sigma + \frac{\alpha H}{2 \mu} + \frac{F}{m^{2}}  \right) \gamma (\phi) + \frac{1}{4 \mu} \theta (\phi) \right\} d \phi,
\end{equation}
\begin{equation}\label{56}
  \mathcal{Y}_{n} =  \frac{1}{16 \pi G} \int_{0}^{2 \pi} e^{in\phi} \left\{ \left( \sigma + \frac{\alpha H}{ \mu} + \frac{F}{m^{2}}  \right) \gamma (\phi) \theta (\phi) + \frac{\theta (\phi)^{2}}{4 \mu} + \frac{\gamma (\phi)^{2}}{\mu l^{2}} \right\} d \phi .
\end{equation}
For the BTZ black hole, we have \cite{5}
\begin{equation}\label{57}
  \gamma  = r_{+}, \hspace{0.7 cm} \theta = \frac{2 r_{-}}{l}, \hspace{0.7 cm} \kappa= \frac{r_{+}^{2} - r_{-}^{2}}{l^{2} r_{+}},
\end{equation}
where $r_{-}$ and $r_{+}$ are inner and outer horizon radiuses, respectively. Thus, the algebra spanned by $\mathcal{T}_{n} $ and $\mathcal{Y}_{n} $ reduced to the following
\begin{equation}\label{58}
  \begin{split}
        i  \left[ \mathcal{T} _{m} , \mathcal{T}_{n} \right] = & - \frac{\kappa ^{2} n}{8 \mu G} \delta _{m+n,0},\\
        i  \left[ \mathcal{Y} _{m} , \mathcal{Y}_{n} \right] = & (m-n) \mathcal{Y}_{m+n},\\
        i  \left[ \mathcal{Y}_{m} , \mathcal{T}_{n} \right] = & - n \mathcal{T}_{m+n} + \frac{\kappa n}{8 \mu G} \left( \frac{ r_{-}}{l} + \alpha H r_{+} \right) \delta _{m+n,0}.
  \end{split}
\end{equation}
where we made a shift on spectrum of $\mathcal{T}_{n} $ by a constant,
\begin{equation}\label{59}
  - \frac{\kappa}{8 \mu G} \left( \frac{ r_{-}}{l} + \alpha H r_{+} \right),
\end{equation}
which is suitable for the following discussion. In this case, Eq.\eqref{55} and Eq.\eqref{56} reduce to
\begin{equation}\label{60}
  \mathcal{T}_{n} = - \frac{\kappa}{4 G} \left\{ \left( \sigma + \frac{\alpha H}{ \mu} + \frac{F}{m^{2}}  \right) r_{+} + \frac{r_{-}}{ \mu l} \right\} \delta _{n,0},
\end{equation}
\begin{equation}\label{61}
  \mathcal{Y}_{n} =  \frac{1}{8 G} \left\{ \left( \sigma + \frac{\alpha H}{ \mu} + \frac{F}{m^{2}}  \right) \frac{2 r_{+} r_{-}}{l} + \frac{(r_{+}^{2} + r_{-}^{2})}{\mu l^{2}} \right\} \delta _{n,0}.
\end{equation}
 Using the results of the paper \cite{10}, we see that the zero mode charge $\mathcal{T}_{0}$ is proportional to the entropy of the BTZ black hole solution of GMMG, i.e. $\mathcal{T}_{0} = \frac{\kappa}{2 \pi} S$, where $S$ is the entropy of the BTZ black hole
\begin{equation}\label{62}
  S = - \frac{\pi}{2 G} \left\{ \left( \sigma + \frac{\alpha H}{ \mu} + \frac{F}{m^{2}}  \right) r_{+} + \frac{r_{-}}{ \mu l} \right\},
\end{equation}
which can be obtained by the formalism presented in \cite{4}, and $\mathcal{Y}_{0}$ gives us the angular momentum of the BTZ black hole, i.e. $j=\mathcal{Y}_{0}$.
\section{Conclusion}
In this paper we have considered the GMMG model in first order formalism. The Lagrangian of the GMMG is given by Eq.\eqref{1}. We have expressed the equations of motion of the GMMG in terms of ordinary torsion-free spin-connection, see equations \eqref{10}-\eqref{13}. Then, we have shown that all the solutions of the Einstein-Hilbert gravity with negative cosmological constant are solutions of the GMMG when equations \eqref{16}-\eqref{18} satisfy. The presence of the Lorentz Chern-Simons term in the Lagrangian makes this model to be non-covariant in the meaning of Lorentz-covariance. We have defined the total variation of dreibein and spin-connection by Eq.\eqref{21} and Eq.\eqref{22}, respectively. Since this model is not Lorentz covariant, so we have used the method introduced in \cite{4,11} to obtain an off-shell conserved current \eqref{29} associated with an arbitrary Killing vector field $\xi$. Thus, by virtue of the Poincare lemma, we have defined the off-shell conserved charge \eqref{36} associated with an arbitrary Killing vector field $\xi$. In section 4, we have reviewed  some results of the paper \cite{5} briefly. The near horizon geometry of a 3D black hole in the Gaussian null coordinates has given by the metric \eqref{39} and components of the metric tensor obey the fall-off conditions \eqref{40} close to the horizon. Also, the near horizon Killing vectors are given by Eq.\eqref{41} and in Fourier modes obey the algebra \eqref{46}. In section 5, we have found the conserved charge \eqref{50} corresponding to the Killing vector \eqref{41}. By introducing Fourier modes of conserved charge, $\mathcal{T}_{n} = Q(e^{in\phi},0) $ and $\mathcal{Y}_{n} = Q(0,e^{in\phi}) $, we have shown that $\mathcal{T}_{n}$ and $\mathcal{Y}_{n}$ obey the algebra \eqref{54}. Then, we have considered the BTZ black hole solution as an example. We have found that in this case the algebra \eqref{54} reduced to \eqref{58}, where we made a shift on spectrum of $\mathcal{T}_{n}$. We saw that the charge associated with time translations, $\mathcal{T}_{0}$, is the product of Hawking temperature $ T_{H}$ and entropy of black hole $S$ , i.e. $\mathcal{T}_{0}= T_{H} S$, also the charge
associated with rotations along $\mathcal{Y}_{0}$ coincides exactly
with the angular momentum of the black hole, i.e. $j=\mathcal{Y}_{0}$. In the limit of $\mu \rightarrow \infty$ and $m \rightarrow \infty$ and by setting $\sigma = -1$, where the GMMG model reduce to the Einstein gravity in the presence of negative cosmological constant, all the results of this paper reduced to the results that have been obtained in \cite{5}. We have shown that in the algebra of Fourier modes of conserved charge, $\mathcal{T}_{n} = Q(e^{in\phi},0) $ and $\mathcal{Y}_{n} = Q(0,e^{in\phi}) $, the central extension term comes from just Lorentz
Chern-Simons term of the GMMG model. But the presence of term proportional to the $\frac{1}{m^2}$ (the term of the NMG), does not lead to a centrally extended algebra. Therefore we can claim that the near horizon
geometry of a non-extremal black hole solutions of NMG, similar to the same solutions of the Einstein gravity, have $BMS_3$ symmetry which can be recovered by means a Sugawara construction.
\section{Acknowledgments}
M. R. Setare thank Gaston Giribet and Hernán González for helpful comments and discussions.


\begin{thebibliography}{9}
\bibitem{5} L. Donnay, G. Giribet, H. A. Gonzalez, M. Pino, Phys. Rev. Lett. \textbf{116} (2016) 091101.
\bibitem{4'}S. Deser, R. Jackiw and S. Templeton, Annals Phys.
140, 372 (1982) [Erratum-ibid. 185, 406.1988 APNYA,
281,409 (1988 APNYA,281,409-449.2000)].
\bibitem{2'} E. A. Bergshoeff, O. Hohm, P. K. Townsend, Phys. Rev. Lett. 102,  201301 (2009).
\bibitem{3'}E. Bergshoeff, O. Hohm, W. Merbis, A. J. Routh and P. K. Townsend,  Class. Quant. Grav. 31, 145008, (2014).
\bibitem{18}A. S. Arvanitakis, A. J. Routh and P. K. Townsend,  Class. Quant. Grav. 31 235012, (2014); A. Baykal, Class. Quant. Grav. 32 (2015) 025013; A. S. Arvanitakis, P. K. Townsend, Class. Quant. Grav. 32 (2015) 085003; B. Tekin, Phys. Rev. D 90, 081701 (2014); M. Alishahiha, M. M. Qaemmaqami, A. Naseh, A. Shirzad, JHEP1412(2014)033 ; G. Giribet, Y. Vásquez, Phys. Rev. D 91, 024026, (2015); M. R. Setare, H. Adami, 	Phys. Rev. D 91, 104039 (2015); A. S. Arvanitakis, Class. Quant. Grav. 32 115010, (2015); M. R. Setare, H. Adami, Gen. Rel. Grav, DOI 10.1007/s10714-016-2033-6, (2016).
\bibitem{4} M.R. Setare, H. Adami, Nucl. Phys. B \textbf{902} (2016) 115.
\bibitem{10} M. R. Setare, H. Adami, Phys. Lett. B \textbf{744} (2015) 280.
\bibitem{2} M. R. Setare, Nucl. Phys. B \textbf{898}, 259 (2015).
\bibitem{5'}H. Bondi, M. van der Burg, and A. Metzner, Proc. Roy.
Soc. London A269, 21 (1962); R. Sachs, Phys. Rev. 128, 2851 (1962).
\bibitem{6'}S. W. Hawking, M. J. Perry, and A. Strominger, (2016),
arXiv:1601.00921 [hep-th].
\bibitem{7'}M. Blau and M. O'Loughlin, arXiv:1512.02858 [hep-th]; R. F. Penna, arXiv:1508.06577 [hep-th]; G. t. Hooft, arXiv:1601.03447 [gr-qc];
M. Bianchi and A. L. Guerrieri, arXiv:1601.03457 [hep-th]; A. Averin, G. Dvali, C. Gomez, and D. Lust, arXiv:1601.03725 [hep-th]; G. Comp`ere and J. Long, arXiv:1601.04958 [hep-th]; A. Kehagias and A. Riotto, arXiv:1602.02653 [hep-th]; G. Comp`ere and J. Long, arXiv:1602.05197 [gr-qc]; H. Afshar, S. Detournay, D. Grumiller, W. Merbis, A. Perez, D. Tempo, R. Troncoso, arXiv:1603.04824 [hep-th].
\bibitem{1} E. A. Bergshoeff, O. Hohm, W. Merbis, A. J. Routh, P. K. Townsend, Lect. Notes. Phys. \textbf{892} (2015) 181.
\bibitem{3} T. Jacobson, A. Mohd, Phys. Rev. D \textbf{92} (2015) 124010.
\bibitem{6} G. Barnich and C. Troessaert, JHEP \textbf{1005} (2010) 062.
\bibitem{7} T. Jacobson, A. Mohd, Phys. Rev. D \textbf{92} (2015) 124010.
\bibitem{8} M. R. Setare, H. Adami, arXiv:1604.07837 [hep-th].
\bibitem{9} G. Barnich, F. Brandt, Nucl. Phys. B \textbf{633} (2002) 3.
\bibitem{19}G. Barnich and G. Compère, Class. Quant. Grav. 24 (2007).
\bibitem{11} Y. Tachikawa, Class. Quant. Grav. \textbf{24} (2007) 737.
\end{thebibliography}
\end{document}